# Nonlinear suppression of dispersion broadening of ultrashort spin-wave pulses in thin YIG films

K. O. Nikolaev[1#], D. Raskhodchikov[2#], J. Bensmann[2#], I. V. Borisenko[1], E. Lomonte[2], L. Jin[2,3], R. Schmidt[2], J. Kern[2], S. Michaelis de Vasconcellos[2], R. Bratschitsch[2], S. O. Demokritov[1], W. H. P. Pernice[2,3], and V. E. Demidov[1]*

[1]*Institute of Applied Physics, University of Muenster, 48149 Muenster, Germany*

[2]*Institute of Physics, Center for Nanotechnology and Center for Soft Nanoscience, University of Muenster, 48149 Muenster, Germany*

[3]*Kirchhoff-Institute for Physics, Heidelberg University, 69120 Heidelberg, Germany*

We study experimentally the nonlinear propagation of short pulses of forward volume spin waves in nanometer-thick YIG films. We show that nonlinearity of the spin system can efficiently counteract dispersion broadening of the pulses, leading to the formation of envelope solitons. We demonstrate that in microscopic YIG systems, microwave powers of the order of one milliwatt are sufficient to reach the soliton formation threshold. At powers slightly above this threshold, we achieve transmission of 3-ns spin-wave pulses over distances of up to 50 micrometers without increase in their temporal width. Our results demonstrate a promising way towards high-rate transmission of information in microscopic spin-wave circuits unaffected by detrimental dispersion effects.

[#]These authors contributed equally

*Corresponding author, e-mail: demidov@uni-muenster.de

# I. INTRODUCTION

Spin waves in magnetic films have long been known as a unique system demonstrating a wide variety of nonlinear dynamic phenomena that have large potential for technical applications in generation, transmission, and processing of high-frequency signals [1-3]. One of the most fascinating and technologically important manifestations of dynamic nonlinearity is its ability to counteract dispersion effects, which are known to strongly limit the information transmission rates in electronic and optical devices and circuits. Indeed, in high-speed systems where information is transmitted in the form of ultra-short pulses, the effects of dispersion broadening are critical even in the case of moderate dispersion, as it occurs in optical fibers [4]. In contrast to light, spin waves in thin films exhibit a relatively strong dispersion. Therefore, the transmission of short spin-wave pulses usually suffers from strong broadening during propagation unless sophisticated phase manipulation or dispersion management techniques are used (see, e.g., [5-11]).

An elegant way to overcome dispersion broadening is to exploit the interplay between dispersion and nonlinearity. Under certain conditions, this interplay is known to lead to the formation of the so-called envelope solitons [4,12] – nonlinear pulses, whose dispersion broadening is completely compensated by nonlinearity. In the past, magnetic envelope solitons have been intensively studied in micrometer-thick magnetic films of Yttrium Iron Garnet (YIG) (see, e.g., [13-18]). It is important to note that envelope solitons strictly retain their shape during propagation only in systems with negligible damping. Since the temporal width of a soliton pulse is directly related to its amplitude, a decrease in the latter due to damping inevitably results in a pulse broadening [4]. The strength of this broadening depends on the ratio of the characteristic spatial scales characterizing the nonlinearity, dispersion, and damping. In typical experiments with micrometer-thick YIG films, dispersion and nonlinearity manifest themself on characteristic spatial scales in the millimeter range. Since spin waves exhibit significant damping at this scale, even in the soliton propagation regime, spin-wave pulses were found to show significant broadening [16].

Recent progress in the preparation of high-quality nanometer-thick films of YIG [19-21] has enabled the development of spin-wave devices and systems with microscopic dimensions [22-25]. Although magnetic damping in such films is typically larger than in previously studied micrometer-thick films, its effects can be less pronounced at microscopic-scale propagation distances. Additionally, due to the peculiarities of the strongly quantized spectrum of magnetic excitation in thin films and nanostructures, it is possible to easily achieve large amplitudes of the magnetization precession and observe nonlinear phenomena that could not be observed in macroscopic-scale systems (see, e.g., [26-30]).

Here we use micro-focus Brillouin light scattering (BLS) spectroscopy to analyze the nonlinear propagation of nanosecond spin-wave pulses in nanometer-thick YIG films. We find that in the low-amplitude regime, the propagation of pulses is accompanied by a more than twofold increase in their temporal width at a propagation distance of 50 μm due to dispersion. We show that, using excitation powers in the milliwatt range, one can drive the spin system into a strongly nonlinear state, where the nonlinearity efficiently counteracts dispersion broadening. Already at a power below one milliwatt, the characteristic nonlinear length becomes comparable to the dispersion length, which corresponds to the condition for the formation of envelope solitons. In this regime, the broadening of the pulses becomes almost suppressed. Moreover, by increasing the excitation power slightly above this threshold, we achieve transmission of spin-wave pulses over distances of 50 μm without any increase in their width. We additionally perform micromagnetic simulations and show that their results quantitatively reproduce our experimental findings.

## II. EXPERIMENT

Figure 1(a) shows a schematic overview of the experiment. We study the propagation of spin-wave pulses in a 110-nm thick film of YIG grown by liquid phase epitaxy on a gadolinium gallium garnet (GGG) substrate. The film is magnetized perpendicular to its plane by a static magnetic field $\mu_0 H_\perp$=300 mT to support the propagation of forward volume spin waves (FVSW) [3]. These waves exhibit a strong positive nonlinear frequency shift [28,29,31] leading to highly efficient self-phase modulation [28]. The latter nonlinear mechanism is known to counteract dispersion effects and lead to the formation of envelope solitons [4]. Additionally, we apply a small in-plane field $\mu_0 H_\parallel$=20 mT to slightly deflect the direction of the static magnetization away from the normal to the film plane, which is necessary to enable magneto-optical detection of FVSW. Note that such a deflection does not significantly modify the dispersion relation of spin waves and does not lead to weakening of the nonlinear frequency shift. We excite spin-wave pulses using a 1-µm wide and 150-nm thick Au antenna fabricated by electron-beam lithography and subsequent lift-off processing, carrying pulses of microwave current. We use an ultra-fast microwave switch with a rise-/fall-time of about 1.5 ns, which allows us to excite spin-wave pulses with a duration as short as 3 ns. The repetition period of the pulses is set to 300 ns, which prevents heating of the YIG film by microwaves.

We image the propagation of pulses with spatial and temporal resolution using micro-focus Brillouin light scattering (BLS) spectroscopy [32]. We focus the probing laser light with a wavelength of 473 nm and a power of 0.25 mW into a diffraction-limited spot [Fig. 1(a)] using a microscope objective lens with a magnification of 100 and a numerical aperture of 0.9 and measure the intensity of light inelastically scattered from spin waves. The intensity of the scattered light (BLS intensity) is proportional to the intensity of the spin waves at the position of the focal spot, which provides high spatial resolution. Temporal resolution is achieved by measuring the delay for each detected scattered photon relative to the moment of time when the excitation pulse is applied to the input antenna.

Figure 1(b) shows dispersion curves of FVSW in the studied film. Symbols show experimental data obtained at low excitation power of $P = 0.1$ mW using BLS operating in the phase-resolved regime [32]. The curves in Fig. 1(b) show the dispersion relation calculated using the micromagnetic simulation package Mumax3 [33] and the method described in detail in [34]. This method allows for an accurate calculation of dispersion curves for different precession angles, which facilitates the analysis of nonlinear phenomena. In these simulations, we use the standard YIG saturation magnetization $\mu_0 M_0 = 175$ mT and exchange constant 3.66 pJ/m.

The dashed curve in Fig. 1(b) shows the dispersion curve calculated for a precession angle of $\theta = 0.1°$, which corresponds to the linear propagation regime. As seen from Fig. 1(b), the calculated curve is in excellent agreement with the experimental data. This fact allows us to use the results of micromagnetic simulations to obtain important dispersion and nonlinear parameters that are difficult to determine experimentally with the required accuracy. In particular, by calculating the first and second derivatives of the dispersion relation, we find the group velocity of the spin waves $v_g = 2\pi \partial f / \partial k$ and the dispersion parameter $D = 2\pi \partial^2 f / \partial k^2$, where $f$ is the frequency and $k$ is the wavenumber. To characterize the nonlinear frequency shift, we calculate the dispersion curve at an increased precession angle of $\theta = 10°$ [solid curve in Fig. 1(b)]. As expected for FVSW, this curve is shifted towards higher frequencies [31]. These data allow us to estimate the nonlinear coefficient $N = 2\pi \partial f / \partial |u|^2$, where $|u| = \frac{m}{\sqrt{2}M_0}$ is the dimensionless amplitude of spin waves, $m$ is the amplitude of the dynamic magnetization, and $M_0$ is the saturation magnetization [31]. We choose $f = 3.85$ GHz as the working frequency for our experiments. It is about 0.15 GHz above the lower cut-off frequency of the spin waves [see Fig. 1(b)], which ensures that the entire frequency spectrum of the 3-ns long pulse falls within the spin-wave frequency band. For this frequency, we obtain $v_g = 0.75$ μm/ns, $D = -0.06$ μm$^2$/ns, and $N = 31$ ns$^{-1}$. We emphasize that $D$ and $N$ have opposite signs. Under these conditions, nonlinearity counteracts dispersion, which is necessary for the suppression of dispersion broadening and formation of envelope solitons [4,35].

## III. RESULTS AND DISCUSSION

Figure 2 shows representative experimental data demonstrating the effects of dispersion and nonlinearity on the propagation of short spin-wave pulses. Here, we plot the temporal profiles of pulses recorded at different distances $x$ from the antenna, as labelled. Experimental data are shown as dots and the curves represent their fits by a Gaussian function. The data in Fig. 2(a) are obtained at low excitation power $P = 0.1$ mW, corresponding to the linear propagation regime, whereas the data in Fig. 2(b) characterize strongly nonlinear propagation at large excitation power $P = 1.5$ mW. As seen from Fig. 2(a), in the linear propagation regime, the pulse exhibits strong dispersion broadening. Its width increases by more than a factor of two at a propagation distance of 50 μm, while the peak intensity decreases by about a factor of five. In contrast, in the nonlinear regime [Fig. 2(b)], the width of the pulse remains almost constant, and its peak intensity decreases significantly slower (by about a factor of 2.5 at a 50-μm distance). This result clearly shows that nonlinearity indeed efficiently counteracts dispersion broadening.

Figure 3 characterizes the observed phenomenon in more detail. In Fig. 3(a) we plot the spatial dependence of the temporal width of the pulse, $w$. The data are obtained by measuring the temporal profiles of pulses at different distances from the antenna with a step size of 2 μm and determining the full width at half maximum (FWHM) of their Gaussian fits. In the linear propagation regime ($P = 0.1$ mW), the width increases monotonically from 3 ns at $x = 0$ to about 6.5 ns at $x = 50$ μm, i.e. by a factor of 2.2. For a pulse with a Gaussian shape, the broadening factor due to dispersion can be described as [4]:

$$\xi = \frac{w}{w_0} = \sqrt{1 + \left(\frac{x}{L_D}\right)^2} \qquad (1)$$

Here, $x$ is the propagation distance, $L_D = \frac{T_0^2 v_g^3}{|D|}$ is the dispersion length, and $T_0 = \frac{w_0}{2\sqrt{\ln(2)}}$, where $w_0$ is the FWHM of the Gaussian intensity profile at $x = 0$. Using the values of the parameters estimated from the analysis of the calculated dispersion curves ($v_g = 0.75$ μm/ns, $D = -0.06$ μm$^2$/ns) and $w_0 = 3$ ns, we obtain $L_D = 23$ μm and $\xi = 2.4$ at $x = 50$ μm, which is very close to the experimental value $\xi = 2.2$. This confirms that the behavior observed at $P = 0.1$ mW is dominated by dispersion effects.

As seen from the data of Fig. 3(a), increasing the input power $P$ from 0.1 to 1.5 mW drastically changes the spatial dependence of the pulse width. In particular, at $P = 1.5$ mW, the pulse exhibits a slight compression at the initial propagation stage ($x = 0$-14 μm) followed by stabilization at $x > 30$ μm at a value close to the initial width $w_0 = 3$ ns. In other words, under these conditions, the nonlinearity completely compensates the dispersion broadening.

Since in the linear propagation regime dispersion broadening also causes the peak intensity of the pulse to decrease at a rate larger than that determined by damping, the intensity of the nonlinear pulse at $P = 1.5$ mW decays significantly slower than the intensity of the linear pulse at $P = 0.1$ mW [Fig. 3(b)]. In other words, the use of the nonlinear propagation regime not only allows for higher data transmission rates due to the absence of broadening, but can also provide the opportunity to significantly improve the signal-to-noise ratio in spin-wave devices operating with short pulses.

Figure 3(c) shows the spatial dependences of the intensity integrated over the duration of the pulse, obtained at $P = 0.1$ and 1.5 mW. In contrast to the data in Fig. 3(b), the integral intensity exhibits a well-defined exponential spatial decay at a rate, which is almost the same at both input powers. This rate is determined solely by the spin-wave damping and corresponds to the decay length of spin waves of $L_A$ = 120 µm or the Gilbert damping parameter of $\alpha$ = $2.4\times10^{-4}$, which is a typical value for high-quality nanometer-thick YIG films [19-21]. This important result indicates that the damping does not increase in the nonlinear propagation regime, as it is usually the case for spin waves in in-plane magnetized films [36-38]. We associate the absence of detrimental nonlinear damping with the vanishing ellipticity of magnetization precession in FVSW, which is known to result in a suppression of nonlinear scattering processes [39].

We now thoroughly analyze the power dependence of the observed nonlinear compensation of dispersion broadening and discuss its relation to the formation of envelope solitons. Figure 4(a) shows the power dependences of the peak intensity of pulses detected at $x$ = 0 and 50 µm. As seen from these data, close to the antenna ($x$ = 0), the peak intensity scales almost linearly with $P$ over the entire range of 0.1-1.5 mW. This observation indicates that the process of excitation of spin-wave pulses by the antenna is mostly unaffected by nonlinearity and that the observed nonlinear behaviors arise at the propagation stage. In contrast, the dependence of the peak intensity detected at $x$ = 50 µm is strongly nonlinear and gives the impression that spin-wave transmission coefficient increases with increasing $P$. As discussed above, this increase is not due to a decrease in spin-wave damping but is caused by a suppression of pulse broadening.

In Fig. 4(b), we show by open triangles the experimental power dependences of the width of the pulses measured at $x$ = 0 and 50 µm. The initial width ($x$ = 0) remains constant over the entire range $P$ = 0.1-1.5 mW. This behavior confirms the above conclusion that nonlinearity does not have a significant effect on the excitation process. In contrast to the initial width, the width of the pulses at a distance $x$ = 50 µm shows a strong power dependence. It decreases monotonically with increasing $P$ and becomes approximately equal to the initial width of 3 ns at $P = 1.5$ mW.

To gain better insight into the observed phenomena, we perform numerical simulations of nonlinear pulse propagation. In accordance with the experimental procedure, we excite spin-wave pulses with a FWHM of 3 ns by applying a local in-plane dynamic magnetic field at a frequency of 3.85 GHz modulated by a Gaussian function and analyze the modification of the pulse width during propagation for different amplitudes of the excitation field. The solid curve in Fig. 4(b) shows the width of the pulses after propagation for 50 μm as a function of the dimensionless intensity $|u|^2 = \frac{m^2}{2M_0^2}$ (upper horizontal scale). As seen from these data, the curve obtained from simulations shows good quantitative agreement with the experimental data. This fact allows us to determine the proportionality coefficient between the dimensionless spin-wave intensity and the microwave power used in the experiment $|u|^2/P = 3.75 \times 10^{-3}$ mW$^{-1}$. Based on the value found, we can estimate the magnetization precession angle $\theta$ achieved in the experiment at different powers. For instance, the smallest power $P = 0.1$ mW corresponds to $\theta = 1.6°$, while the largest power $P = 1.5$ mW corresponds to $\theta = 6.1°$.

The obtained calibration data also allow us to estimate the threshold power at which the formation of envelope solitons is expected. The widely used criterion for the formation of solitons is the equality of the dispersion length $L_D$ and the so-called nonlinear length $L_{NL} = \frac{\pi v_g}{|N||u|^2}$, which is dependent on the pulse intensity [4]. Using the above-determined values $L_D = 23$ μm, $v_g = 0.75$ μm/ns, and $N = 31$ ns$^{-1}$, we obtain the threshold intensity $|u|^2 = 3.3 \times 10^{-3}$, which corresponds to $P \approx 0.9$ mW. This value is marked in Fig. 4(b) by a dashed vertical line. We note that under these conditions the spin-wave pulse still experiences a slight broadening of about 20%. We associate this residual broadening with effects of the damping, as discussed above. Strictly speaking, in a system with finite damping it is not possible to achieve a situation where a nonlinear pulse exactly retains its shape during propagation. However, by exciting pulses with an intensity slightly exceeding the soliton threshold, one can realize a situation where the pulse width after a certain propagation distance is approximately equal to the initial width. In fact, this is achieved in the experiment at $P = 1.5$ mW [see Fig. 3(a)]. Under these conditions, at the initial propagation stage ($x = 0$-$14$ μm), the large intensity of the pulse results in dominant nonlinear compression, leading to a decrease in the pulse width. As the intensity decreases due to damping, the effects of nonlinearity weaken and the pulse starts to experience dispersion broadening. The two effects cannot exactly compensate each other at every point in space. Nevertheless, compensation on average can be achieved. We emphasize that the deviations of the pulse width from its initial value during the propagation depend on the strength of the damping. In our system, the decay length $L_A = 120$ μm is much larger than the dispersion length $L_D = 23$ μm and the nonlinear length. Therefore, the deviations of the pulse width are minimal.

## IV. CONCLUSIONS

In this study, we have shown experimentally that nonlinearity of the spin system in out-of-plane magnetized YIG films of nanometer thickness makes it possible to transmit short spin-wave pulses without detrimental dispersion broadening. Thanks to the strong nonlinearity and the absence of nonlinear damping, this can be achieved at low microwave powers suitable for technological applications. The possibility to overcome dispersion broadening can be used to implement high-speed magnonic integrated circuits that require high-rate transmission and processing of information. Additionally, the demonstrated nonlinear effects may be useful for the implementation of unconventional computing schemes using nonlinear spin-wave phenomena.

# ACKNOWLEDGMENTS

This work was supported by the Deutsche Forschungsgemeinschaft (DFG, German Research Foundation) – project number 529812702 and 433682494 – SFB 1459.

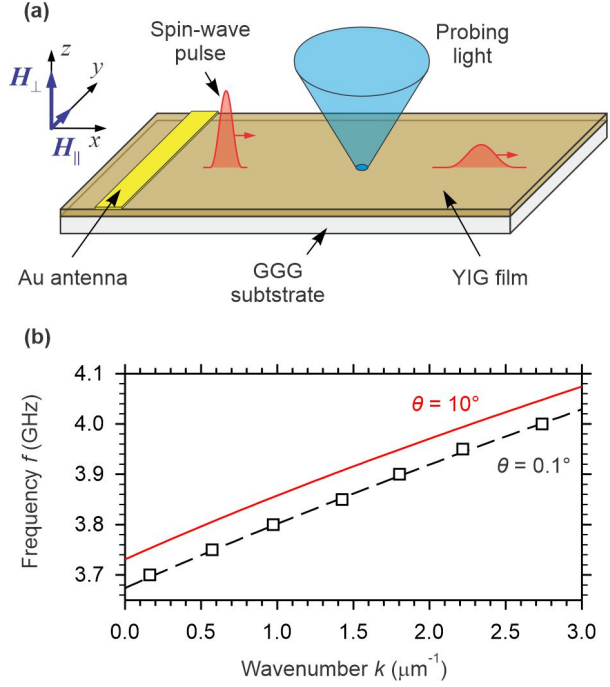

FIG. 1. (a) Schematics overview of the experiment. 3-ns long spin-wave pulses propagate in a 110-nm thick YIG film. The film is magnetized perpendicular to its plane by a static magnetic field $\mu_0 H_\perp$=300 mT. An additional small in-plane field $\mu_0 H_\parallel$=20 mT is applied to enable magneto-optical detection of spin waves. The propagation of spin-wave pulses is analyzed using space- and time-resolved micro-focus BLS spectroscopy. (b) Dispersion curves of spin waves. Symbols show experimental data. Curves show the results of micromagnetic simulations. Solid and dashed curves are calculated at two different precession angles $\theta$, as labelled.

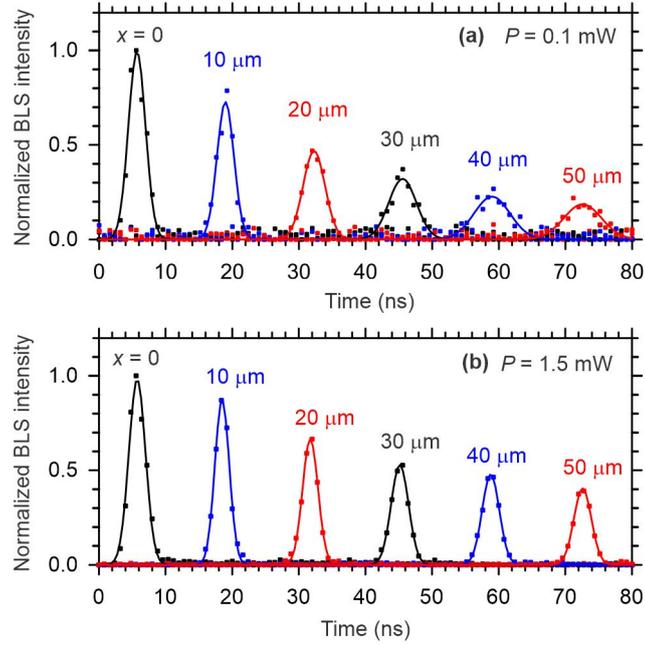

FIG. 2. Temporal profiles of pulses recorded at different distances $x$ from the antenna, as labelled. Experimental data are shown as dots and the curves represent their fits by a Gaussian function. The data in (a) are obtained at low excitation power $P = 0.1$ mW corresponding to the linear propagation regime. The data in (b) characterize strongly nonlinear propagation at large excitation power $P = 1.5$ mW.

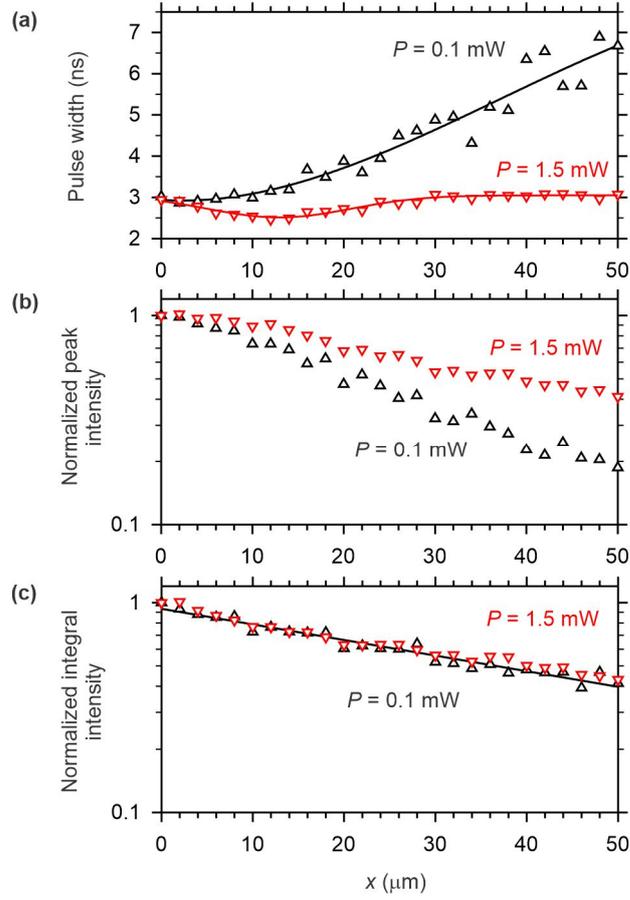

FIG. 3. (a) Spatial dependence of the temporal width of spin-wave pulses. (b) Spatial dependence of the normalized peak intensity. (c) Spatial dependence of intensity integrated over the duration of the pulse. The data are obtained at $P = 0.1$ and 1.5 mW, as labelled. Symbols show the experimental data. Curves in (a) are guides to the eye. The curve in (c) shows the exponential fit of the experimental data.

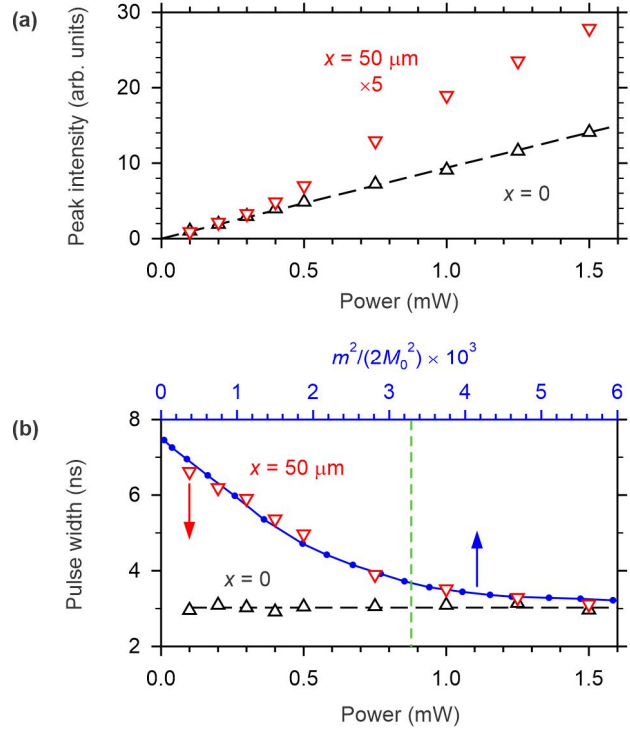

FIG. 4. (a) Power dependences of the peak intensity of pulses detected at $x = 0$ and 50 μm, as labelled. Symbols show experimental data. The dashed line is a linear fit of the data at $x = 0$. The data at $x = 50$ μm are multiplied by a factor of 5 for clarity. (b) Triangles show the experimental power dependences of the temporal width of spin-wave pulses detected at $x = 0$ and 50 μm, as labelled. The dashed horizontal line marks the mean value at $x = 0$. Solid dots show the dependence of the temporal width of spin-wave pulses after a propagation distance of 50 μm on their dimensionless intensity (upper horizontal scale), obtained from micromagnetic simulations. The solid curve is a guide to the eye. The vertical dashed line marks the threshold of soliton formation.